\documentclass[aps,floatfix,a4paper,showpacs,twocolumn,nofootinbib,superscriptaddress,10pt]{revtex4}
\usepackage{graphicx,float}\usepackage{graphicx,float}
\usepackage[all]{xy}
\usepackage{amsmath}
\usepackage{amssymb}
\usepackage{color}
\usepackage{epsfig}		
\usepackage{graphicx,epstopdf}
\usepackage{subfigure}
\usepackage{pdfpages}
\usepackage[colorlinks,hyperindex]{hyperref}

\hypersetup{
colorlinks,
citecolor=blue,
linkcolor=black,
urlcolor=black,
}

\newcommand{\bes}{\begin{subequations}}
\newcommand{\ees}{\end{subequations}}
\def\ben{\begin{eqnarray}}
\def\een{\end{eqnarray}}
\def\be{\begin{eqnarray}}
\def\ee{\end{eqnarray}}

\usepackage[colorlinks,hyperindex]{hyperref}

\hypersetup{hidelinks,backref=true,pagebackref=true,hyperindex=true,colorlinks=true,breaklinks=true,urlcolor= blue}
\hypersetup{%
  colorlinks = true,
  linkcolor  = blue
}

\begin{document}

\title{AdS/QCD duality and the quarkonia holographic information entropy}
\author{Nelson R. F. Braga}
\affiliation{Instituto de F\'isica, Universidade Federal do Rio de Janeiro,
Caixa Postal 68528, RJ 21941-972 - Brazil}
\email{braga@if.ufrj.br}
\author{Rold\~ao da Rocha}
\affiliation{CMCC, Universidade Federal do ABC, UFABC, 09210-580, Santo Andr\'e, Brazil}
\email{roldao.rocha@ufabc.edu.br}
\begin{abstract}

Bottomonium and charmonium, representing quarkonia states, are scrutinized under {\color{black}{the point of view of}} the  information theory,  in the AdS/QCD holographic  setup.
A logarithmic measure of information, comprised by the configurational entropy, is here employed to quantitatively study quarkonia radially excited $S$-wave  states.  
The configurational entropy provides data regarding the relative dominance and the {\color{black}{abundance}} of the bottomonium and charmonium states, whose underlying information is  more compressed, in the Shannon's theory meaning. The derived 
{\color{black}{configurational}} entropy, therefore,  identifies the lower phenomenological prevalence of higher $S$-wave resonances and higher masses quarkonia in Nature.
\end{abstract}
\maketitle


\section{Introduction}

The {\color{black}{configurational}} (CE) entropy is a paradigm, contemporarily proposed on the beginning of this decade \cite{Gleiser:2011di,Gleiser:2012tu}, that considers foreseeable aspects of information into general systems. The {\color{black}{configurational}} entropy, as a logarithmic quantity of information, measures the form  intricacy of localized, Lebesgue-integrable, physical configurations of general systems \cite{Gleiser:2014ipa,Sowinski:2015cfa,Gleiser:2012tu}. 
The  CE, which directly measures the compression of lossless data into the frequency modes of a system, is derived from the physical system configuration via a Fourier transform.  It expresses the quantity of the organizational capacity enciphered into the  energy density describing a physical system, as the temporal component of the energy-momentum tensor of the theory. Beyond the principle of least action, where physical systems extremize the corresponding action, the CE,   {\color{black}{in addition, points to either more stable states, or to more abundant/dominant ones}}.  Physical states with higher CE either require more energy to be produced in Nature, or are less observed/detected than their CE-stable peer states, or even both. 
The lattice approach of Shannon information entropy has the  information dimension, driving data distributions,  
as an upper bound for the compression data rate of any variable in a distribution. Statistical-mechanics grounds were introduced in this context, in the literature  \cite{Bernardini:2016hvx}.  The CE encodes the precise measurement of information that is mandatory to account  the spatial shape of Lebesgue-integrable functions, with respect to their parametrical arguments. 

The CE has been used  to derive the abundance of light-flavor mesons in AdS/QCD holographic models, both  in the asymptotically AdS$_5$ metric of $D_3$ brane systems  and in the Sakai--Sugimoto $D_4$-$D_8$ system, as well, in Ref.  
 \cite{Bernardini:2016hvx}. Besides, glueball  phenomenology, their stability and the relative dominance of their states, according to their spin, were   investigated in the context of the CE, in the  AdS/QCD setup 
 \cite{Bernardini:2016qit}. The CE was also used in QCD models to study and refine cross-sections of hadrons in the color-glass condensate \cite{Karapetyan:2016fai,Karapetyan:2017edu}. Moreover, the CE was employed to study AdS-Schwarzschild black holes \cite{Braga:2016wzx,Lee:2017ero}, in a consistent setup that corroborates to the  consistency of the Hawking-Page phase transition \cite{Braga:2016wzx}, and also to study compact stars  
 \cite{Gleiser:2013mga,Gleiser:2015rwa}, including the prediction of the Chandrasekhar critical density of 
 a Bose--Einstein condensate of gravitons into compact stars,  on fluid branes \cite{Casadio:2016aum}.  Solitons in a cold quark-gluon plasma were further investigated with the CE, in Ref.  \cite{daSilva:2017jay}, to derive the pulse soliton width for which the soliton spatial profile has more compressed information, matching phenomenological data. 
 In addition, the CE was successfully utilized in particle physics to determine the Higgs boson mass \cite{Alves:2017ljt} and  to derive the precise mass of an axion that  interacts with neutrinos and photons \cite{Alves:2014ksa}. 
Other topological defects were also scrutinized in the context of the CE \cite{Correa:2015vka,Correa:2016pgr}.

The fundamental theory of strong interactions, quantum chromodynamics (QCD),  presents a coupling constant that increases as the energy decreases. This is the reason why there are many important hadronic properties that can not be perturbatively described by QCD. 
An interesting complementary tool for studying hadronic states in the vacuum or at finite temperature and (or) density are the AdS/QCD models.
This approach is motivated by the AdS/CFT correspondence \cite{Maldacena:1997re,Gubser:1998bc,Witten:1998qj},  
relating weakly coupled supergravity in five dimensional  anti-de Sitter space (AdS$_5$) to a conformal super Yang-Mills theory on the boundary. 
Phenomenological AdS/QCD models assume the existence of a similar type of duality in cases where  conformal invariance is broken by the introduction of an energy parameter in the AdS background, representing an infrared (IR) cut-off -- mass scale  --  in the gauge theory.  
The simplest one is the hard wall model,  proposed in Refs. \cite{Polchinski:2001tt,BoschiFilho:2002ta,BoschiFilho:2002vd}, that consists in placing a hard cut-off in anti-de Sitter (AdS)  space.   
Another AdS/QCD model is the soft wall one, that has the  property that the square of the mass linearly grows  with the radial excitation number \cite{Karch:2006pv}. 
In this case, the background involves an AdS  space and a scalar field that effectively acts as a smooth  infrared cut-off. A review of AdS/QCD models can be  found in Ref.  \cite{Brodsky:2014yha}. 

Quarkonia  are hadronic states made of a heavy quark-antiquark pair. For the vectorial case, they are
the charmonium $J/\psi$ meson, composed by a charmed $\bar{c}c$ pair, the bottomonium $\Upsilon$ meson,  made of a $\bar{b}b$ pair, and their corresponding radial excitations. 
Due to  the high mass of the top quark, toponium does not exist. In fact, the top quark decays through the electroweak interaction before a bound state can form, providing a rare example of a weak process proceeding more quickly than a strong process. Usually in the literature, the word quarkonium refers only to charmonium and bottomonium, and not to any of the lighter quark-antiquark states \cite{Bernardini:2003ht}.

A consistent AdS/QCD model for charmonium and bottomonium in the vacuum was recently proposed  in \cite{Braga:2015jca},  extended to finite temperature in  \cite{Braga:2016wkm}   and to finite density and temperature in \cite{Braga:2017oqw}. 
There are two different energy parameters in this model. One infrared mass scale and one large ultraviolet scale, related to the non hadronic decay of the heavy vector mesons. The model of Ref. \cite{Braga:2015jca} leads to decay constants that decrease with the radial excitation number of the meson states. Such a behavior is experimentally  observed and was not reproduced by the standard previous AdS/QCD models.  
Here we will develop the calculation of the CE for heavy vector mesons, using this model that provides consistent spectra of masses and decay constants.  

This work is organized as follows: in Sect. II, the CE is briefly reviewed, and Sect. III is devoted to review  the AdS/QCD holographic model for heavy vector mesons,  relating the decay constants to  the two-point correlation function. Then, in Sect. IV, the holographic model for quarkonia is employed to calculate the CE for bottomonium and charmonium states, as a function of the radial excitation level, where scaling relations can be observed.  Our final conclusions are drawn in Sect. V.

\section{Configurational entropy and Shannon information entropy}
 
The CE shall be here applied to study the entropic information content of quarkonia, within the AdS/QCD setup. 
The CE, based on the Shannon's information theory, comprises a procedure that logarithmically measures the underlying information of quadratically Lebesgue-integrable functions, denoted hereon by $\rho(x)$ to further represent the energy density that underlies the system to be analyzed,  defined on $\mathbb{R}^d$. The Fourier transform 
\begin{eqnarray}
\rho(k) = \int_{\mathbb{R}^d}d^dx\; \rho(x)e^{-ik\cdot x},\label{ftrans}
\end{eqnarray} is the main ingredient to construct the modal fraction 
\cite{Gleiser:2012tu}
\begin{eqnarray}\label{modall}
\check\rho(k) = \frac{|\rho(k)|^{2}}{\int_{\mathbb{R}^d} d^dk |\rho(k)|^{2}},\label{modalf}
\end{eqnarray}
which represents the weight of every single mode tagged by $k$. Hence the CE, associated with the sistem energy density,  is given by \cite{Gleiser:2012tu} 
\begin{eqnarray}
S[\rho] = - \int_{\mathbb{R}^d} d^dk\,\check{\rho}(k)\log \check{\rho}(k).\label{ce1}
\label{ce}
\end{eqnarray}
 The underlying  informational features in Eq. (\ref{ce}) are apparent by its analogy to the entropy defined by Shannon, $
S=- \sum_{k} p_{k} \log p_{k},$ 
where the $\{p_{j}\}$ denotes a set of probability densities inherent to any data set ${A}_1$, manifesting into a string encrypted in another ${A}_2$ data set \cite{Gleiser:2012tu, Bernardini:2016hvx,daSilva:2017jay}. The quantity of information that is necessary to decipher the 
regarded string relies upon the code employed. The less information is necessary for the data decodification, the more compressed  the information itself may be. The intensity of the information compression is yielded by the Shannon's entropy. Therefore, considering the CE, the profile of a spatially bounded, Lebesgue-integrable, function $\rho(x)$ 
emulates a string in the original data set, it being the case that its Fourier transform $\rho(k)$ represents the profile encoded into the momentum data set, labelled by $\{k\}$. In addition, the modal fraction in Eq. (\ref{modall}) regards  the share of the configuration, in momentum space, to the  profile of $\rho(x)$. Hence,  the CE represents the inherent informational content in general physical systems that have (localized) energy density configurations $\rho(x)$.

In the next section, we describe the holographic model that describes the $S$-wave radial resonances of quarkonia.

\section{Holographic model for quarkonia}    

Thermal properties of heavy vector mesons into a plasma of quarks and gluons are useful instruments for investigating heavy ion collisions \cite{Matsui:1986dk}.  
The dissociation of quarkonia states inside a plasma was described in the AdS/QCD holographic framework in Ref. \cite{Braga:2016wkm} (see also \cite{Braga:2017oqw,Braga:2017bml}). 
The first radial excitations $1S, 2S$ and $ 3S$ were revealed as spectral function peaks, whose height increase according to the decrement of the temperature.  This description was obtained by extending to finite temperature the model proposed in  \cite{Braga:2015jca},   for decay constants and masses of heavy vector mesons. 

Vector mesons are represented, in the soft wall model \cite{Karch:2006pv}, by a vector field  $V_m = (V_\mu,V_z)\,$ ($\mu = 0,1,2,3$) living in AdS space, 
assumed to be dual to the gauge theory current
$J^\mu(x) = \bar{q}(x)\gamma^\mu q(x)$, for quark spinor fields $q(x)$. The action for the vector field reads: 
\begin{equation}
I \,=\, \int dz\,d^4x \, \sqrt{-g} \,\, e^{-\Phi (z)} \, \left(- \frac{1}{4 g_5^2} F_{mn} F^{mn}
\,  \right) \,\,, 
\label{vectorfieldaction}
\end{equation}
\noindent where $F_{mn} = \partial_mV_n - \partial_n V_m$ and $\Phi(z) = \kappa^2z^2   $ 
denotes the soft wall dilaton background, responsible for the IR cut-off. The energy parameter  $\kappa$ is   related to the mass scale. The anti-de Sitter space AdS$_5$ is endowed  with a metric   
\begin{equation}\label{mmetric}
 ds^2 =g_{mn}dx^mdx^n= e^{2{\rm A}(z)}(-dt^2 + d\vec{x}\cdot d\vec{x} + dz^2)\,,
\end{equation}
 with warp factor ${\rm A}(z) = -\log(z/R)$, for $z\in (0,\infty)$.
With the gauge choice  $V_z=0$, the boundary value of the vector field $V^0_{\mu}(x) =
\lim_{z\to 0} V_\mu (x,z)$ spacetime components play the role of sources of the correlation functions for the boundary current operator  $  J^\mu (x)$ 
\begin{eqnarray}
\!\! \langle 0 \vert \, J_\mu (x) J_\nu (x') \,  \vert 0 \rangle= \frac{\delta}{\delta V^{0}_{\mu}(x)} \frac{\delta}{\delta V^{0}_{\nu}(x')}
e^{- I_{\rm on\, shell}},
 \label{2pointfunction}
\end{eqnarray}
\noindent where $ I_{\rm on\, shell}$  denotes the on shell action, obtained by the boundary term: 
 \begin{equation}
I_{\rm on \, shell }\,=\, - \frac{1}{2 {\tilde g}_5^2}  \, \int d^4x \,\, \frac{e^{- \kappa^2 z^ 2  } V^\mu \partial_z V_\mu }{z}
\Big\vert_{z \to 0}
 \,.
\label{onshellaction}
\end{equation}
The effective dimensionless coupling is {\color{black}{driven by}} ${\tilde g}_5^2 =  \frac{g_5^2}{R}$. Since we want to describe particle states, one goes to momentum space with respect to the coordinates $x^\mu$ and {\color{black}{split}} the vector field as 
 \begin{equation} 
 V_\mu (p,z) \,=\, v (p,z) V^0_\mu ( p ) \,,
 \label{Bulktoboundary}
\end{equation}  
 \noindent where    $ v (p,z)   $ is known as the  bulk-to-boundary propagator, satisfying the following equation of motion:
\begin{eqnarray}
\left(\partial_z^2-2\kappa^2z^2\partial_z + p^ 2\right) v (p,z)= 0\,.\nonumber
\label{BulktoboundaryEOM}
\end{eqnarray}
The boundary condition $v (p, z)\vert_{z\to0} = 1$ must be imposed in such a way that $V^0_\mu ( p )$ plays the role of the source of the gauge theory currents correlators. 

In momentum space, the so-called two-point function $\Pi (p^2) $ is associated with the current-current correlator   by {\color{black}{the following expression}}: 
\begin{eqnarray}
\! \!\! \left(p^2 \eta_{\mu\nu} \!-\!  p_\mu p_\nu   \right) \Pi (p^2) 
\!=\!\!\int\! d^4x' e^{-ip\cdot x'} \langle 0 \vert \, J_\mu (x') J_\nu (0) \,  \vert 0 \rangle, 
\label{correlatorand2pointfunction}
\end{eqnarray} 
where $\eta_{\mu\nu}$ denotes the boundary metric components. 
So, the two-point can be holographically represented as
 \begin{eqnarray}
  \Pi ( p^2 )  =   \frac{1}{{\tilde g}_5^ 2 \, (-p^ 2) }  \frac{ e^{ -\kappa^2 z^2 }}{z}  \,  v (p,z) \partial_z v (p,z) 
\Big\vert_{z \to 0} \,. \label{hol2point}
\end{eqnarray}      
 On the other hand, the two-point function possess a spectral decomposition, with respect to the masses, $m_n$, and to the states decay constants,  $f_n$, {\color{black}{given by}}:
  \begin{equation}
\Pi (p^2)  = \sum_{n=1}^\infty \, \frac{f_n^ 2}{(- p^ 2) - m_n^ 2 + i \epsilon} \,. 
\label{2point}
\end{equation} 
As discussed in Refs. \cite{Karch:2006pv,Braga:2015jca}, 
 the soft wall model yields the masses and decay constants, respectively, as  
 \begin{eqnarray}
 m_n &=& 4 \kappa^2 n \,,\qquad f_n   \, = \, 
  \frac{\sqrt{ 2}\kappa}{{\tilde g}_5}\,,
 \label{decayconstants}
\end{eqnarray}  
\noindent where $n=1,2,3,\ldots\, $. This means that, in the soft wall model, all the vector meson radial excitations have the same decay constant. This differs from the result obtained from experimental data that shows a decrement of the decay constants decrease as the radial excitation level $n$ increases. 

 In order to overcome this problem with the decay constants, an alternative holographic model for heavy vector mesons was proposed  in Ref.  \cite{Braga:2015jca}. In contrast to the original soft wall model, that contains only one dimensionfull parameter  $\kappa$, Ref.   \cite{Braga:2015jca} introduces an  additional parameter, by holographically calculating the operator product of currents in Eq. (\ref{2pointfunction}), when the radial coordinate has a finite  location $z = z_0$, whose inverse corresponds to an ultraviolet (UV)  energy  scale. 
  The bulk-to-boundary propagator is then expressed as a solution of Eq. (\ref{BulktoboundaryEOM}), {\color{black}{precluding the range}} $0 < z < z_0$ and  satisfying the new boundary condition, $v(p,z)\vert_{z\to z_0} = 1,$   
 \begin{eqnarray} 
v (p,z ) \, = \, \frac{ U \left(\frac{p^ 2}{4\kappa^ 2}, 0, \kappa^2 z^ 2 \right) }{U \left(\frac{p^ 2}{4\kappa^ 2}, 0, \kappa^2 z_0^ 2 \right)}\,,
 \label{bulktoboundary2}
\end{eqnarray}  
\noindent where $U(a,b,c)$ denotes the Tricomi's confluent hypergeometric function. As the boundary is placed at $z= z_0$, hence the on shell action reads
 \begin{eqnarray}
I_{\rm on \, shell }\,=\, - \frac{1}{2 {\tilde g}_5^2}  \, \int d^4x \,\,\, \frac{e^{- \kappa^2 z^ 2  } V^\mu \partial_z V_ \mu}{z} 
{\big \vert}_{\, z \to z_0} \,,
\label{onshellaction2}
\end{eqnarray}
\noindent and the two-point function has the following expression: 
\begin{equation}
  \Pi ( p^2 )  \, =\,   \frac{1}{{\tilde g}_5^ 2 \, (-p^ 2) }  \frac{ e^{ -\kappa^2 z^2 }}{z}v (p,z) \partial_z v (p,z) 
  {\Big\vert}_{\,z \to z_0} 
   \, .
 \label{hol2pointnew}
\end{equation} 

\noindent     Then, using the bulk-to-boundary propagator in  Eq. (\ref{bulktoboundary2}), and a recursion iteration for the Tricomi's functions one  finds:
 \begin{equation} 
 \Pi ( p^ 2)  \, = \, \frac{1}{2 {\tilde g}_5^ 2 }\frac{  U \left(1 + \frac{p^ 2}{4\kappa^ 2}, 1, \kappa^2 z_0^ 2\right) }{U \left(\frac{p^ 2}{4\kappa^ 2}, 0, \kappa^2 z_0^ 2\right)}e^ {-\kappa^ 2 z_0^ 2}\,.
 \label{correlator2}
\end{equation} 
  In this approach  \cite{Braga:2015jca}, the masses and decay constants are acquired from the analysis of the two-point function poles.  The singularities arise from the zeroes of the denominator in Eq. (\ref{correlator2}).  Hence, the masses of the states at radial excitation level $n$ read  $  m_n^2 = - p_n^2$, with $ U \left(\frac{p_n^ 2}{4\kappa^ 2}, 0, \kappa^2 z_0^ 2\right) \,=\,0$. 
  
Near an open neighbourhood at $p^ 2 = p^ 2_n$, which corresponds to a simple pole, the two-point function can be approximated by 
\begin{equation}
 \Pi (p^2)   \approx \frac{f_n^ 2}{(-p^ 2)  + p_n^ 2 }\Big\vert_{p^ 2 \to p^ 2_n} \,,
\label{2pointlimit}
\end{equation} 
\noindent and one identifies the coefficients of the expression near the pole to the decay constant $f_n$,  by  Eq. (\ref{2point}).   
 In the non hadronic decay, a heavy vector meson annihilates into light leptons. Then, the UV scale has the order of the heavy  meson  masses and the parameter $\kappa$ is identified to the heavy quark mass.    The found decay constants  using this approach decrease with {\color{black}{respect to}} with respect to the excitation level, as observed from experiments.  The coupling ${ \tilde g}^2_5 = g^2_5/R$ can be acquired by {\color{black}{using the analogy to QCD} \cite{Karch:2006pv}, yielding} ${ \tilde g}_5 \, =\, 2 \pi $.
  
In the next section we compute  the CE underlying the charmonium and  bottomonium states using this model. For a similar model for heavy mesons see \cite{Braga:2015lck} and for holographic models for heavy-light mesons see,  for example \cite{Liu:2016iqo,Liu:2016urz}

\section{Configurational entropy of quarkonia}

The localized, quadratically integrable function $\rho(z)$  in Eq. (\ref{ftrans}) corresponds to the energy density of the considered system.   The energy momentum tensor for an action integral of the form $I \,=\, \int d^4x dz \, \sqrt{-g} \,L  $ reads
 \begin{equation}
 T_{mn}(z)=  \frac{2}{\sqrt{ - g }} \left[ \frac{\partial (\sqrt{-g} L)}{\partial g^{mn} } \,
 - \, \frac{\partial}{\partial x^r }  \frac{\partial (\sqrt{-g}  L)}{\partial \left( \frac{\partial g^{mn} }{\partial x^r}\right) }\,
  \right].
  \label{energymomentum}
 \end{equation} 
 \noindent  Quarkonia states are described by the vector fields   $V_\mu$  with action integral   given by Eq. (\ref{vectorfieldaction}). In this case, the  second term of 
  Eq. (\ref{energymomentum}) is absent,  since the action does not depend on derivatives of the metric.  The energy density corresponds to the $T_{00} $ component of the energy momentum tensor, that reads  
\begin{eqnarray}
\rho(z)&=&T_{00}(z)=\frac{  e^{-\kappa^2z^2} }{g^2_5}\, \Big[    \, g_{00}\left(\frac{1}{4}g^{mp}g^{nq}F_{mn} F_{pq}\right)\nonumber\\&&\qquad\qquad-  g^{mn}F_{0n} F_{0m} \Big]  \,.
\end{eqnarray}
Taking a plane wave solution in the meson rest frame $V_\mu=\varepsilon_\mu v(p,z) e^{-imt}$, without loss of generality, with 
$\varepsilon_\mu=(0,1,0,0)$, yields
\begin{eqnarray}
\rho (z)=\frac{z^2e^{-\kappa^2z^2}}{2 g^2_5 R^2}\left(-\left(\partial_z v\right)^2+  m^2 v^2\right) .
\label{density}
\end{eqnarray}
Note that for the plane wave complex solution the appropriate field strength term in the action (\ref{vectorfieldaction})   is $ F^\ast_{mn} F^{mn}  $. 
The solution for $v(p,z) $ is given by Eq. (\ref{bulktoboundary2}). Considering the heavy meson states at radial excitation level $n$ {\color{black}{to be on shell}}, $p^2 = - m_n^2$, the solution reads 
\begin{eqnarray}
v_n(m_n , z) = \frac{U \left(-\frac{m_n^2}{4\kappa^ 2}, 0, \kappa^2 z^2\right)}{U \left(-\frac{m_n^2}{4\kappa^ 2}, 0, \kappa^2 z_0^ 2\right)} \,.
\label{solonshell}
\end{eqnarray}
Denoting by $\kappa_c$ and $\kappa_b$ the  values of the parameter $\kappa$, respectively for charmonium and bottomonium, a suitable fit to the existing experimental data of \cite{Agashe:2014kda} is obtained,  using  $   \kappa_c = 1.2 {\rm GeV},$ $\kappa_b = 3.4 {\rm GeV}$, and $z_0 = 0.08 {\rm GeV}^{-1}$ as in Ref. \cite{Braga:2015jca}. 
We show on tables I and II the masses of the states of charmonium and bottomonium, respectively, obtained from the roots of the algebraic equation   $ U \left(-\frac{m_n^ 2}{4\kappa^ 2}, 0, \kappa^2 z_0^ 2 \right) \,=\,0$. 
 
 In the present case the energy density is a function of just one variable, the coordinate $z$, that is defined in the range $ z_0 \le z < \infty $. So, we can write Eq. (\ref{ftrans}) in the form $ \rho ( k ) = C (k) + i S (k) $ where:
\begin{equation}
\!\!\!\!\!\!\!C(k) \!=\! \int_{z_0}^\infty\!\! \!\rho (z) \cos (kz) dz, \,\, S(k) \!=\! \int_{z_0}^\infty\!\! \!\rho (z) \sin (kz) dz.
\end{equation}
Then our version of Eq. (\ref{modalf}) is taken in the form
\begin{equation}
\check\rho(k) = \frac{ C^2(k) + S^2(k)  }{ \int_0^\infty dk^\prime \left[   C^2( k^\prime ) + S^2 (k^\prime ) \right] },\label{modal2}
\end{equation}
and the configurational entropy is written as
\begin{eqnarray}
S[\rho] = - \int_{0}^\infty dk\,\check{\rho}(k)\log \check{\rho}(k).\label{ce2}
\label{ce2}
\end{eqnarray}
Eqs. (\ref{density}-- \ref{ce2})  are then used together to compute the CE for the quarkonia states, using the appropriate values of the parameters.
Our results are listed in tables I and II.
\begin{table}[h]
\centering
\begin{tabular}[c]{||c|c||c||}
\hline\hline 
\multicolumn{3}{||c||}{Chamonium} \\
\hline\hline
$S$-wave resonance &  Masses (MeV)   & CE \\
\hline\hline
$\,\, 1S \,\,$ & $ 2410.03$ &$\;1.42952\;$ \\ 
\hline
$\,\, 2S \,\,$ & $ 3409.87$ &$\;1.75312\;$ \\ 
\hline 
$\,\, 3S \,\,$ & $ 4174.54$ &$\;1.92768\;$ \\ \hline
$\,\, 4S \,\,$ & $ 4820.13$ &$\;2.05353\;$ \\ \hline
$\,\, 5S \,\,$ & $ 5388.12$ &$\;2.15396\;$ \\ \hline
$\,\, 6S \,\,$ & $ 5901.73$ &$\;2.23796\;$ \\ \hline
$\,\, 7S \,\,$ & $ 6374.14$ &$\;2.31022\;$ \\ \hline
$\,\, 8S \,\,$ & $ 6813.79$ &$\;2.37361\;$ \\ \hline
\hline
\end{tabular}   
\caption{Masses and configurational entropies  of charmonium states as a function of their radial excitation level.   }\medbreak\medbreak\medbreak \medbreak\medbreak\medbreak \medbreak 
\centering
\begin{tabular}[c]{||c|c||c||}
\hline\hline
\multicolumn{3}{||c||}{Bottomonium} \\
\hline\hline
 $S$-wave resonances&  Masses (MeV)   & CE \\
\hline\hline
$\,\, 1S \,\,$ & $ 7011.34$ &$\;2.42517\;$ \\ 
\hline
$\,\, 2S \,\,$ & $ 9883.82$ &$\;2.81458\;$ \\ 
\hline 
$\,\, 3S \,\,$ & $ 12077.60$ &$\;2.98732\;$ \\ \hline
$\,\, 4S \,\,$ & $ 13923.91$ &$\;3.11226\;$ \\ \hline
$\,\, 5S \,\,$ & $ 15546.11$ &$\;3.21205\;$ \\ \hline
$\,\, 6S \,\,$ & $ 17011.72$ &$\;3.29546\;$ \\ \hline
$\,\, 7S \,\,$ & $ 18358.19$ &$\;3.36716\;$ \\ \hline
$\,\, 8S \,\,$ & $ 19610.34$ &$\;3.43002\;$ \\ \hline
\hline
\end{tabular}   
\caption{Masses and configurational entropies  of bottomonium states as a function of their radial excitation level.   }
\end{table}

\begin{figure}[H]
\begin{center}
\includegraphics[width=3.29in]{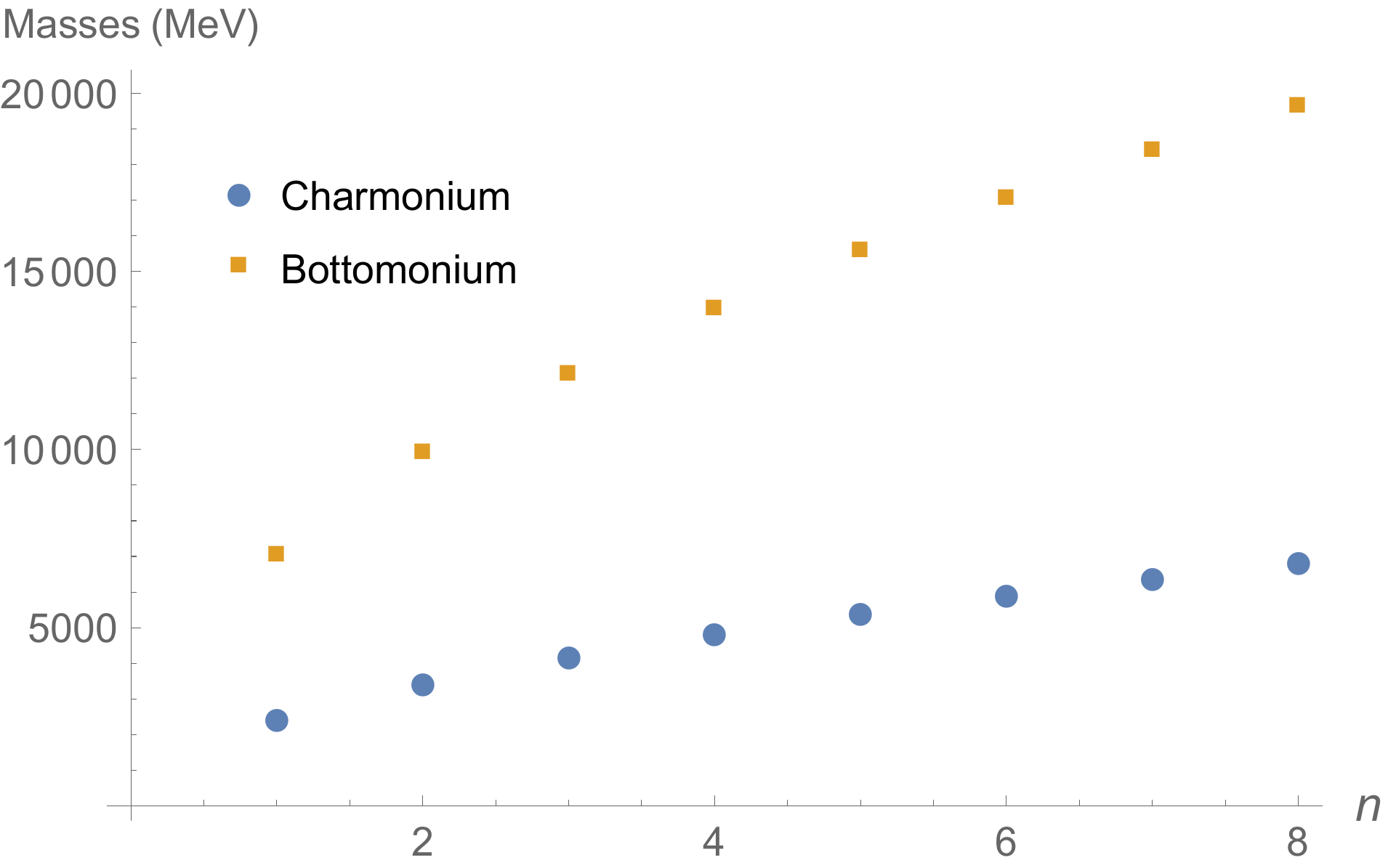}
\caption{\footnotesize Masses of quarkonia states as a function of their radial excitation level. }
\end{center}
\end{figure}

\begin{figure}[H]
\begin{center}
\includegraphics[width=3.19in]{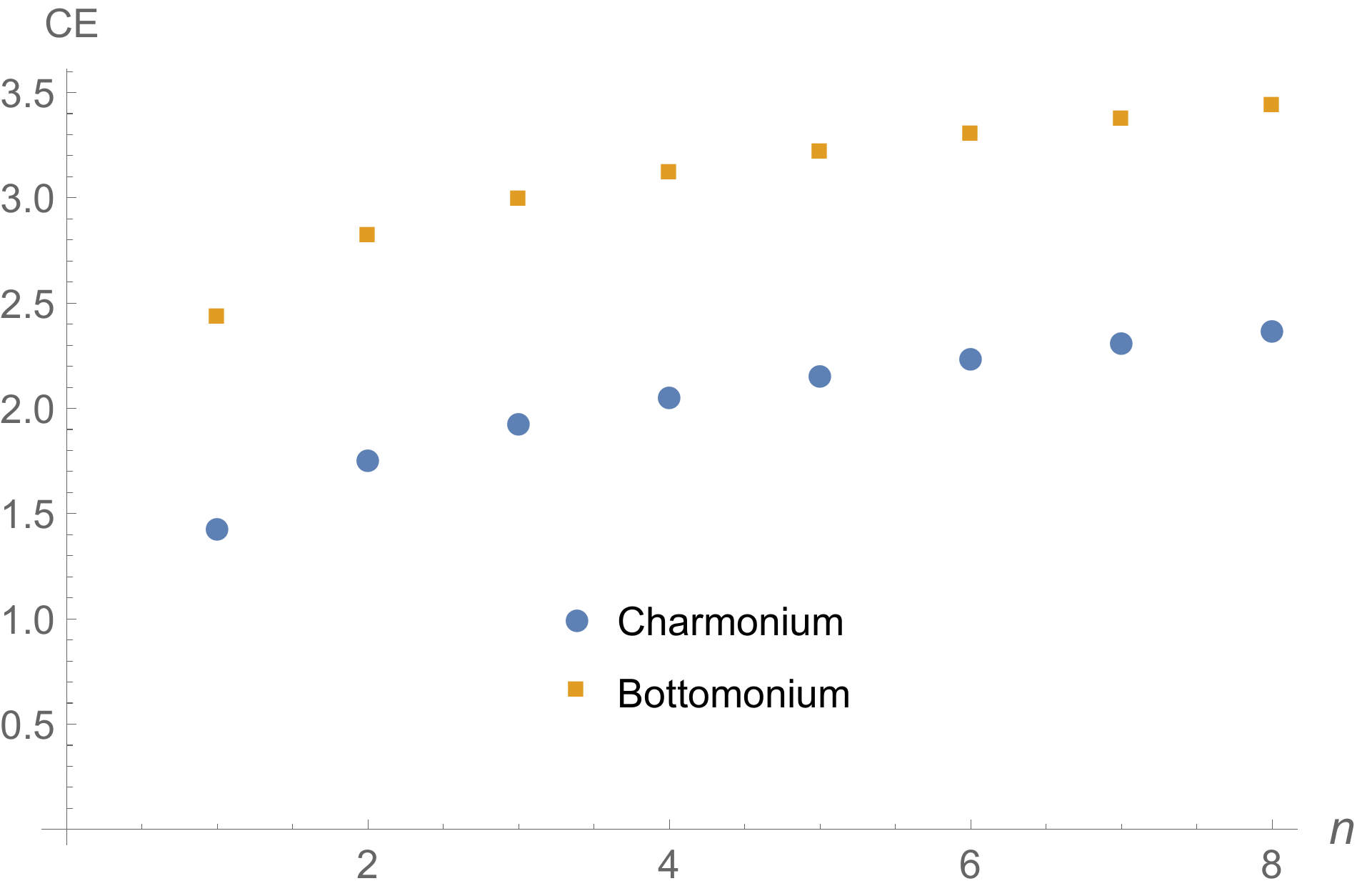}
\caption{\footnotesize Configurational entropy (CE) of quarkonia as a function of the radial excitation level for $S$-wave resonances.}
\end{center}
\end{figure}

\begin{figure}[H]
\begin{center}
\includegraphics[width=3.29in]{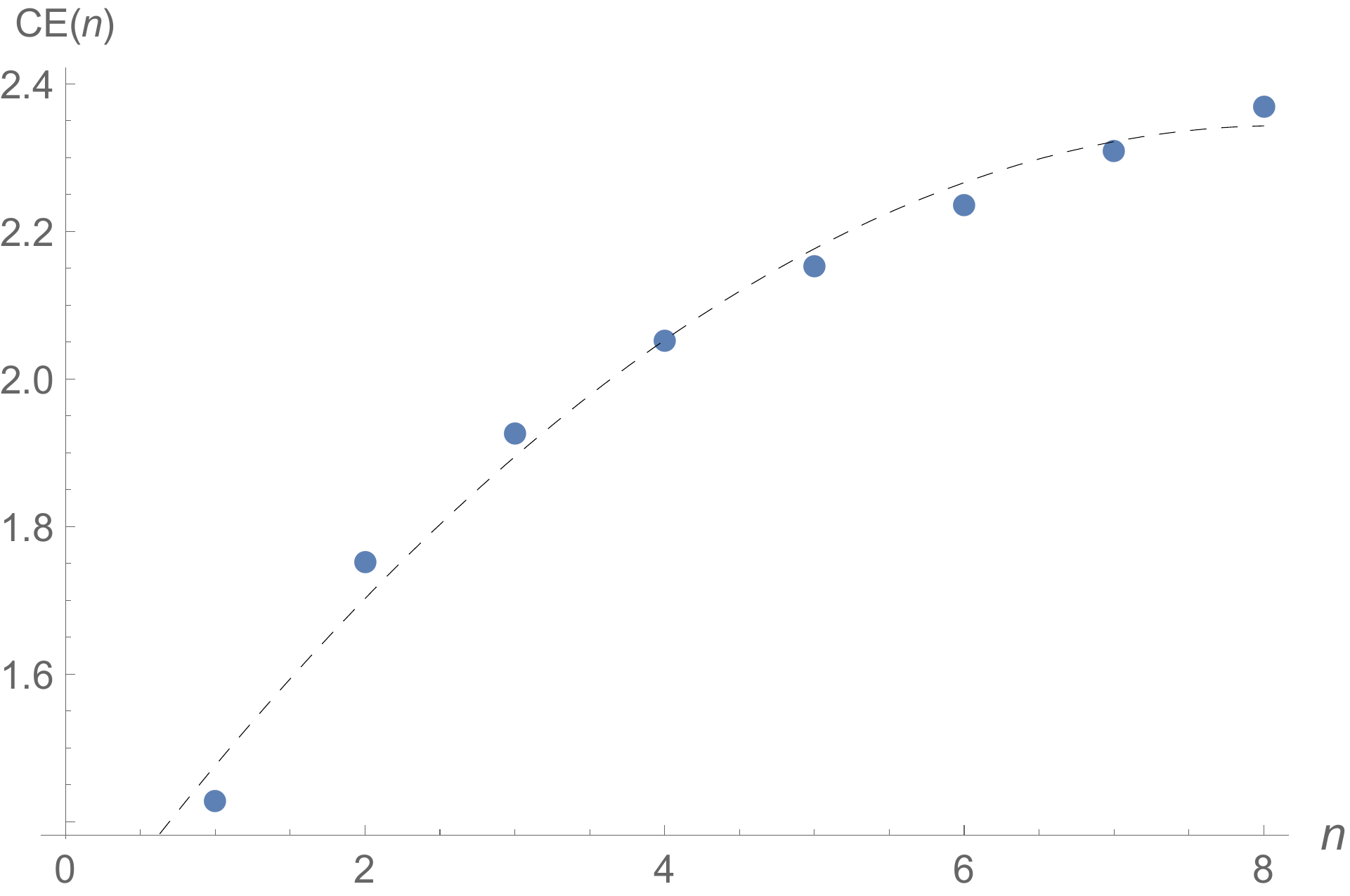}
\caption{\footnotesize Configurational entropy (CE) as a function of the charmonium state radial excitation level (gray list plot); the dashed line represents a quadratic interpolation.}
\end{center}
\end{figure}

\begin{figure}[H]
\begin{center}
\includegraphics[width=3.19in]{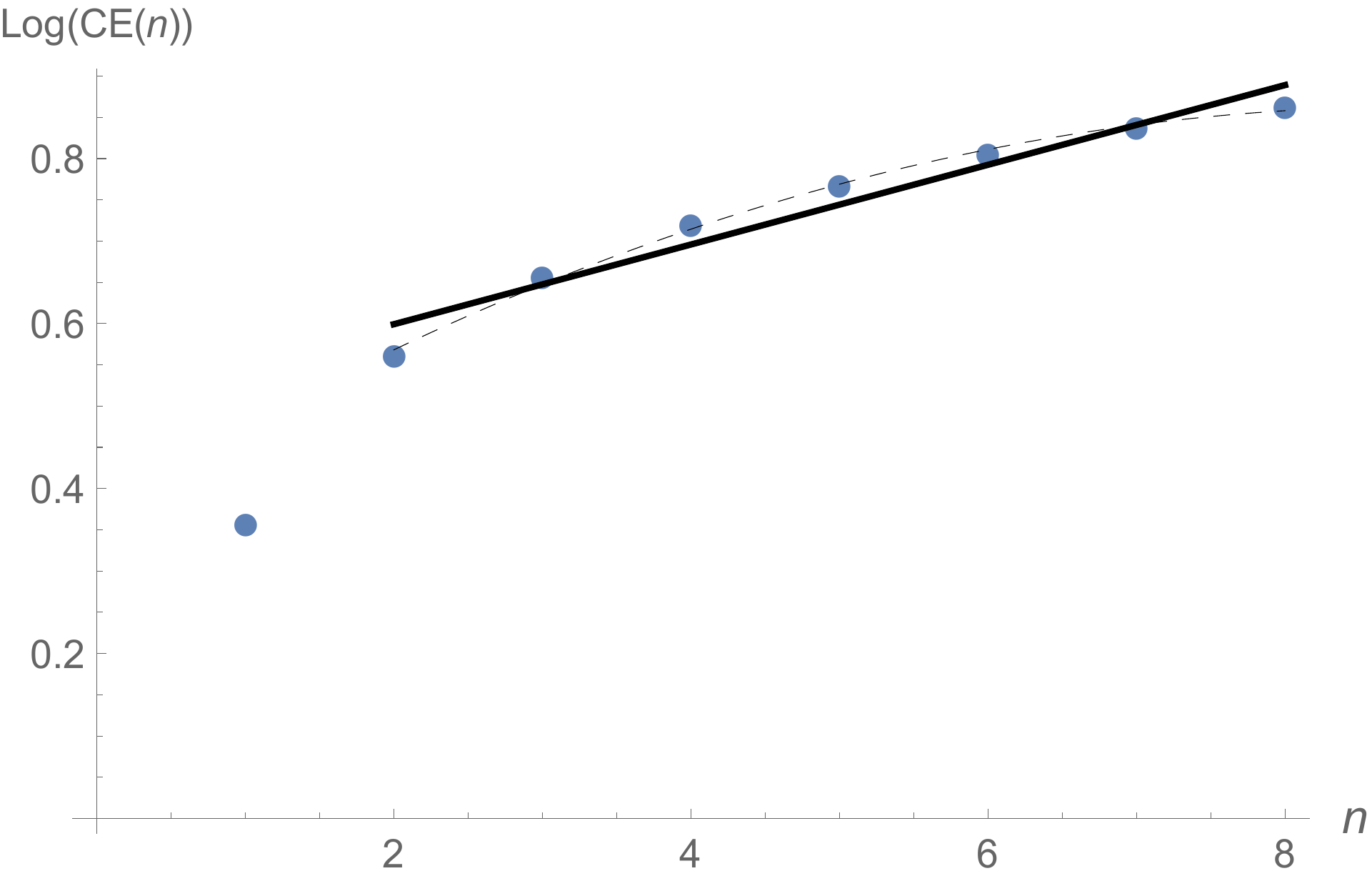}
\caption{\footnotesize Logarithm of the configurational entropy (CE) as a function of the charmonium state radial excitation level  (gray list plot); the black line refers to the linear regression and the dashed line represents a quadratic fit, both accomplished in the range $1<n\leq 8$.}
\end{center}
\end{figure}

\begin{figure}[H]
\begin{center}
\includegraphics[width=3.19in]{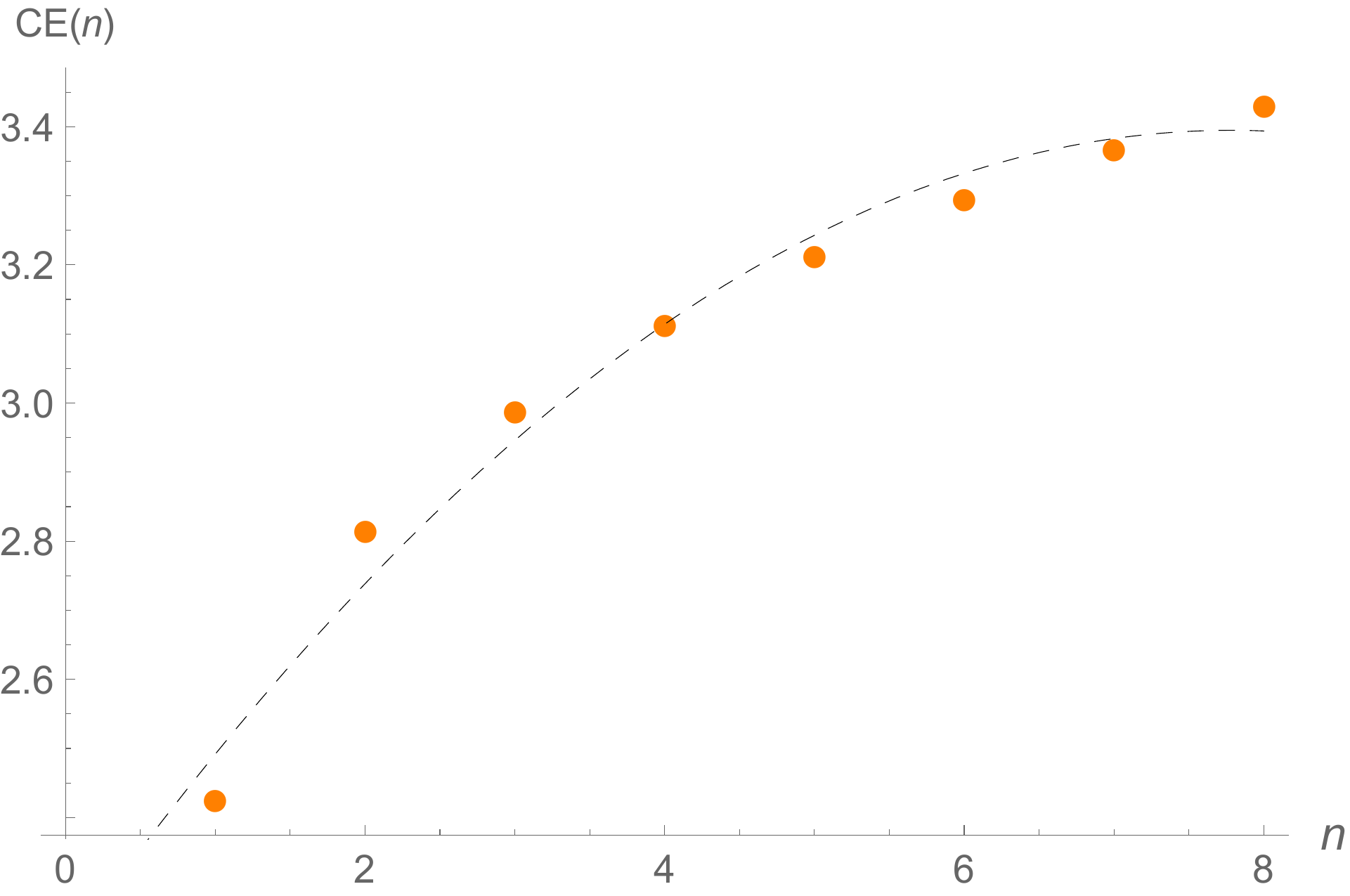}
\caption{\footnotesize Configurational entropy (CE) as a function of the bottomonium state radial excitation level (orange list plot);  the dashed line stands for a quadratic interpolation.}
\end{center}
\end{figure}

\begin{figure}[H]
\begin{center}
\includegraphics[width=3.19in]{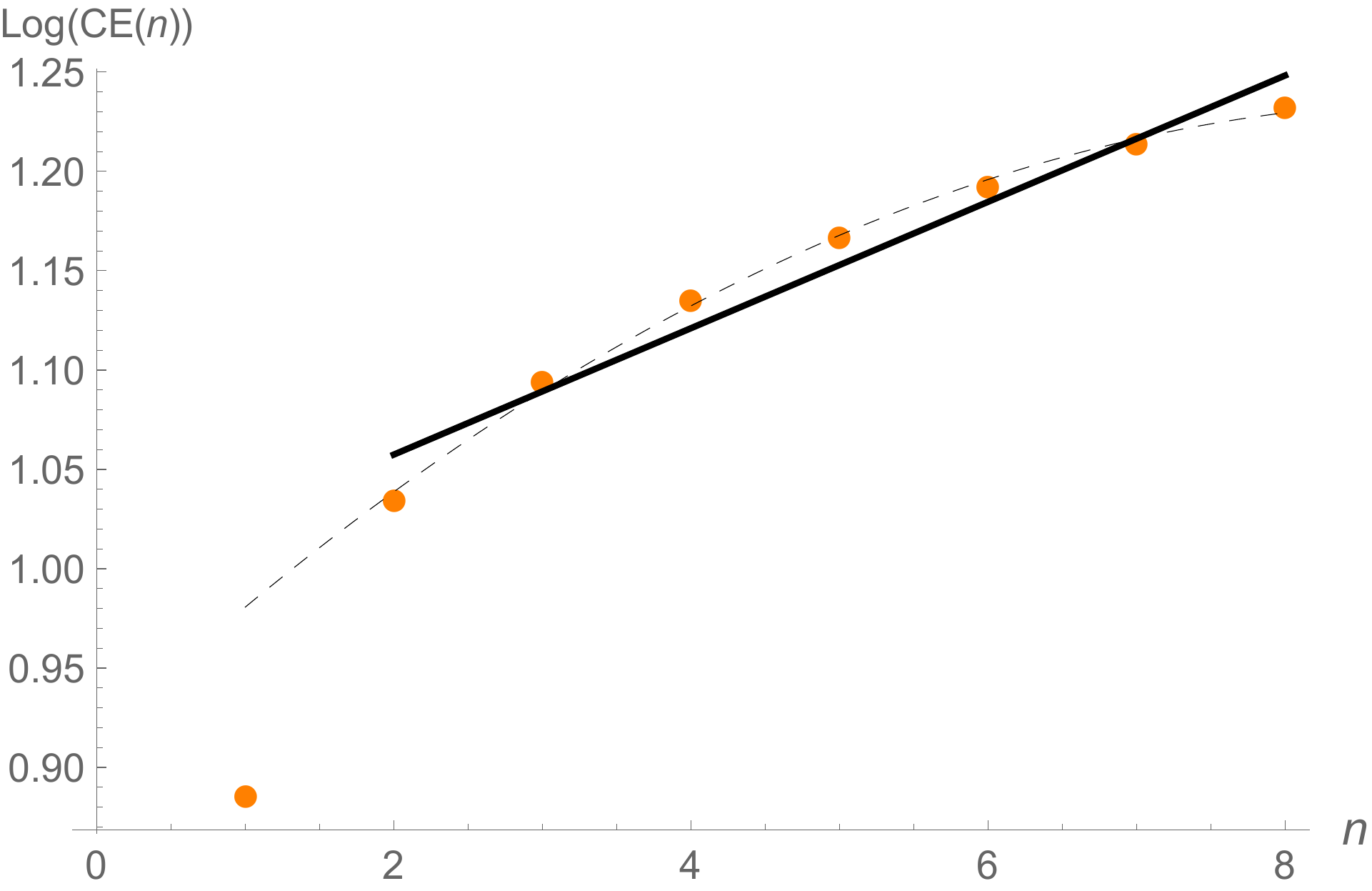}
\caption{\footnotesize Logarithm of the configurational entropy (CE) as a function of the bottomonium state excitation level  (orange list plot); the black line refers to the linear regression and the dashed line represents a quadratic fit, both accomplished in the range $1<n \leq 8$.}
\end{center}
\end{figure}

Figs. 1--6 show that the CE increases as value of the quarkonia radial excitation level increases, indicating that    quarkonia states with  lower excitation level have more dominance.  
By taking  the logarithm of the configurational entropy as a function of the charmonium radial excitation level, the possibility of a scaling relation between the CE and the charmonium excitation level can be derived, which is illustrated in Fig. 4. {\color{black}{
 In fact, the logarithm of the CE can be remarkably approximated by the linear function
 Log(CE($n$)) = $0.0483 n+0.5025$, at least in the range $1<n \leq 8$, when we implement the linear regression procedure as a 
  linear interpolation approach for modelling the relationship between the CE and the charmonium states excitation $n$ level. This linearization method is depicted as the black line in Fig. 4. The involved error 
underlying this linear approach is a tiny one, consisting of 4.1\%. Besides, the logarithm of the CE  can be also quadratically interpolated by the equation  Log(CE($n$)) = $-0.0052 n^2 + 0.1104 n +0.3721$, plot into the dashed line in Fig. 4. The very small value of the quadratic coefficient in this last quadratic polynomial equation yields a quadratic scaling relation between the  logarithm of the CE and charmonium states radial excitation level $n$, with an error of 0.2\%,  in the range $1<n \leq 8$.

Analogously, a similar analysis can be accomplished for the logarithm of the CE, as a function of the bottomonium state radial excitation $n$ level. Indeed, the logarithm of the CE can be linearly approximated by  Log(CE($n$)) = $0.0318 n+0.9938$, with an error of 4.6\%, in the range $1<n\leq 8$. The linearization procedure implemented by the linear regression method in then illustrated by the black line in Fig. 5. Moreover, the logarithm of the CE can be quadratically interpolated by the equation  Log(CE($n$)) = $- 0.0037 n^2+ 0.0693 n+0.9151$, in the range $1<n\leq 8$. This quadratic fit has an error of 0.3\%. Analogously to the charmonium states, the the logarithm of the CE has then a  linear scaling with the bottomonium states radial excitation level.}}

\section{Concluding remarks and outlook}
As discussed in Sects. I and II, recent studies indicate that  the configurational entropy (CE)
provides information about either the relative stability or dominance, among different  states of a physical system. 
Here we have calculated the CE for heavy vector mesons by considering their supergravity holographic AdS/QCD  duals. The results obtained for both bottomonium and charmonium states are that the CE increases with respect to the radial excitation level.
This result is consistent with the observed fact that the higher excited states are in general less produced, or less abundant, that the lower states. Such a result reinforces the expectation of  universality of the role of the CE. 
{\color{black}{
Our results here presented might be further generalized to other cases when the warp factor in
Eq. (\ref{mmetric}) yields non-conformal deformations of the AdS$_5$ metric  \cite{dePaula:2008fp}. Strictly in the context of the CE, this model has been successfully applied, for providing information on the stability of the glueball states, as well as their relative dominance according to their spin, in Ref. \cite{Bernardini:2016qit}. In addition, this model was also used to 
study  light-flavor mesons \cite{Bernardini:2016hvx} and 
might be adapted to encompass the setup developed in Sect. III. Nevertheless, it is still far beyond the proposed theme, regarding our results, and to carefully settle this setup has been revealing an intricate task.}}

\section*{Acknowledgments}
NRFB is partially supported by CNPq  (grant No. 307641/2015-5). RdR is grateful to CNPq (grant No. 303293/2015-2) and  to FAPESP (Grants No.~2015/10270-0 and No. 2017/18897-8), for partial financial support.

\end{document}